\documentclass{article}

\usepackage{PRIMEarxiv}

\usepackage[utf8]{inputenc}
\usepackage[T1]{fontenc}
\usepackage{hyperref}
\usepackage{url}
\usepackage{booktabs}
\usepackage{amsfonts}
\usepackage{nicefrac}
\usepackage{microtype}
\usepackage{graphicx}
\usepackage{listings}
\usepackage{xcolor}
\usepackage{fancyhdr}

\graphicspath{{media/}}

\lstdefinestyle{json}{
    basicstyle=\ttfamily\footnotesize,
    breaklines=true,
    frame=single,
    backgroundcolor=\color{gray!10},
    keywordstyle=\color{blue},
    stringstyle=\color{red},
    commentstyle=\color{green!50!black},
    numbers=left,
    numberstyle=\tiny\color{gray},
    stepnumber=1,
    numbersep=5pt,
}

\title{OpenID Connect for Agents (OIDC-A) 1.0: A Standard Extension for LLM-Based Agent Identity and Authorization
\thanks{\textit{\underline{GitHub Repository}}: \url{https://github.com/subramanya1997/oidc-a/}}
}

\author{
  Subramanya Nagabhushanaradhya \\
  \texttt{subramanyanagabhushan@gmail.com } \\
}

\usepackage{float}
\begin{document}
\maketitle

\begin{abstract}
OpenID Connect for Agents (OIDC-A) 1.0 is an extension to OpenID Connect Core 1.0 that provides a comprehensive framework for representing, authenticating, and authorizing LLM-based agents within the OAuth 2.0 ecosystem. As autonomous AI agents become increasingly prevalent in digital systems, there is a critical need for standardized protocols to establish agent identity, verify agent attestation, represent delegation chains, and enable fine-grained authorization based on agent attributes. This specification defines standard claims, endpoints, and protocols that address these requirements while maintaining compatibility with existing OAuth 2.0 and OpenID Connect infrastructure. The proposed framework introduces mechanisms for agent identity representation, delegation chain validation, attestation verification, and capability-based authorization, providing a foundation for secure and trustworthy agent-to-service interactions in modern distributed systems.
\end{abstract}

\keywords{OpenID Connect \and OAuth 2.0 \and LLM Agents \and Agent Authentication \and Delegation Chains \and Attestation \and Authorization}

\section{Introduction}

\subsection{Rationale}

As Large Language Model (LLM) based agents become increasingly prevalent in digital ecosystems, there is a growing need for standardized methods to represent their identity and manage their authorization. Traditional OAuth 2.0 \cite{rfc6749} and OpenID Connect \cite{openid-connect-core} protocols were designed primarily for human users and conventional applications, lacking the necessary constructs to represent the unique characteristics of autonomous agents.

LLM-based agents differ from traditional applications in several fundamental ways. They act on behalf of users with varying degrees of autonomy, ranging from fully supervised assistants to highly autonomous decision-making systems. They operate within complex delegation chains where authority flows from users through potentially multiple intermediate agents. Their capabilities are dynamic and based on their underlying models, which may evolve over time. Furthermore, they require mechanisms to attest to their integrity, origin, and operational characteristics to establish trust with relying parties.

This specification addresses these gaps by extending OpenID Connect to provide a comprehensive framework for agent identity and authorization. By building upon established OAuth 2.0 and OpenID Connect standards, OIDC-A ensures compatibility with existing infrastructure while introducing the necessary extensions for the agent ecosystem.

\subsection{Terminology}

This specification uses the terms defined in OAuth 2.0 \cite{rfc6749}, OpenID Connect Core 1.0 \cite{openid-connect-core}, and introduces the following additional terms specific to agent-based systems:

\begin{itemize}
\item \textbf{Agent}: An LLM-based software entity capable of autonomous or semi-autonomous action based on natural language instructions and contextual understanding \cite{llm-agents-survey}.
\item \textbf{Agent Provider}: The organization responsible for creating, training, deploying, and/or hosting the agent infrastructure.
\item \textbf{Agent Model}: The specific LLM model that powers the agent (e.g., GPT-4 \cite{gpt4}, Claude 3 \cite{claude3}, Gemini Pro).
\item \textbf{Agent Instance}: A specific running instance of an agent, typically associated with a particular task, conversation, or execution context.
\item \textbf{Delegator}: The entity (typically a human user or another agent) who delegates authority to an agent to act on their behalf.
\item \textbf{Delegation Chain}: A sequence of delegation steps from the original user through potentially multiple intermediate agents, representing the transitive flow of authority.
\item \textbf{Attestation}: Cryptographic proof of an agent's integrity, origin, operational characteristics, and/or compliance with specific policies \cite{rfc9334}.
\item \textbf{Attestation Evidence}: A data structure containing the cryptographic proof and associated metadata used for attestation verification \cite{ietf-rats-eat}.
\item \textbf{Relying Party (RP)}: In this context, a Resource Server or Client application that needs to verify an agent's identity, authorization, and attestation before granting access to protected resources.
\end{itemize}

\subsection{Overview}

OIDC-A extends OpenID Connect through six primary mechanisms:

First, it defines new standard claims for representing agent identity, including agent type, model, version, provider, and instance identifiers. These claims enable relying parties to make informed authorization decisions based on agent characteristics.

Second, it specifies comprehensive mechanisms and formats for agent attestation evidence, allowing agents to prove their integrity and origin using cryptographic techniques compatible with frameworks such as IETF RATS (Remote Attestation Procedures Architecture) \cite{rfc9334} and Entity Attestation Tokens \cite{ietf-rats-eat}.

Third, it establishes protocols for representing and validating delegation chains, enabling transparent tracking of authority flow from original users through potentially multiple agents while enforcing scope reduction and constraint propagation.

Fourth, it provides discovery mechanisms for agent capabilities and attestation support, allowing relying parties to understand what agents can do and how to verify their claims.

Fifth, it defines authorization frameworks suitable for agent-specific use cases, including capability-based access control and constraint enforcement mechanisms.

Finally, it introduces dedicated endpoints for attestation verification and capability discovery, standardizing the interfaces through which relying parties can validate agent claims and discover agent capabilities.

\section{Agent Identity Claims}

\subsection{Core Agent Identity Claims}

The foundation of OIDC-A is a set of standardized claims that represent agent identity and characteristics. These claims extend the standard OpenID Connect claims to capture agent-specific attributes. Table~\ref{tab:core-claims} presents the core agent identity claims that MUST or SHOULD be included in ID Tokens issued to or about agents.

\begin{table}[h]
\caption{Core Agent Identity Claims}
\centering
\begin{tabular}{p{3cm}p{2cm}p{6cm}p{2cm}}
\toprule
\textbf{Claim} & \textbf{Type} & \textbf{Description} & \textbf{Requirement} \\
\midrule
\texttt{agent\_type} & string & Identifies the type/class of agent (e.g., ``assistant'', ``retrieval'', ``coding'') & REQUIRED \\
\texttt{agent\_model} & string & Identifies the specific model (e.g., ``gpt-4'', ``claude-3-opus'') & REQUIRED \\
\texttt{agent\_version} & string & Version identifier of the agent model & RECOMMENDED \\
\texttt{agent\_provider} & string & Organization that provides/hosts the agent & REQUIRED \\
\texttt{agent\_instance\_id} & string & Unique identifier for this specific instance & REQUIRED \\
\bottomrule
\end{tabular}
\label{tab:core-claims}
\end{table}

These core claims provide the fundamental identification information necessary for relying parties to understand what agent they are interacting with. The \texttt{agent\_type} claim allows for classification of agents based on their intended purpose or specialization. The \texttt{agent\_model} and \texttt{agent\_version} claims identify the specific LLM implementation, which is crucial for understanding the agent's capabilities and limitations. The \texttt{agent\_provider} claim establishes accountability by identifying the organization responsible for the agent. Finally, the \texttt{agent\_instance\_id} provides a unique identifier for tracking and auditing purposes.

\subsection{Delegation and Authority Claims}

Delegation is a fundamental concept in agent-based systems, where agents act on behalf of users or other agents. Table~\ref{tab:delegation-claims} describes the claims used to represent delegation relationships and authority transfer.

\begin{table}[h]
\caption{Delegation and Authority Claims}
\centering
\begin{tabular}{p{3.5cm}p{1.5cm}p{6.5cm}p{2cm}}
\toprule
\textbf{Claim} & \textbf{Type} & \textbf{Description} & \textbf{Requirement} \\
\midrule
\texttt{delegator\_sub} & string & Subject identifier of the entity who most recently delegated authority & REQUIRED \\
\texttt{delegation\_chain} & array & Ordered array of delegation steps & OPTIONAL \\
\texttt{delegation\_purpose} & string & Description of the purpose/intent for delegation & RECOMMENDED \\
\texttt{delegation\_constraints} & object & Constraints placed on the agent by the delegator & OPTIONAL \\
\bottomrule
\end{tabular}
\label{tab:delegation-claims}
\end{table}

The \texttt{delegator\_sub} claim provides immediate visibility into who granted authority to the agent, establishing a direct accountability link. The \texttt{delegation\_chain} claim, described in detail in Section~\ref{sec:delegation-chain}, captures the complete history of authority transfer when multiple delegation steps occur. The \texttt{delegation\_purpose} claim documents the intended use case for the delegation, supporting audit and compliance requirements. The \texttt{delegation\_constraints} claim allows delegators to impose restrictions on agent actions, such as time limits, resource restrictions, or operational boundaries.

\subsection{Capability, Trust, and Attestation Claims}

Beyond identity and delegation, agents require claims to communicate their capabilities, trustworthiness, and attestation status. Table~\ref{tab:capability-claims} describes these additional claims.

\begin{table}[h]
\caption{Capability, Trust, and Attestation Claims}
\centering
\begin{tabular}{p{3.5cm}p{1.5cm}p{6.5cm}p{2cm}}
\toprule
\textbf{Claim} & \textbf{Type} & \textbf{Description} & \textbf{Requirement} \\
\midrule
\texttt{agent\_capabilities} & array & Array of capability identifiers & RECOMMENDED \\
\texttt{agent\_trust\_level} & string & Trust classification of the agent & OPTIONAL \\
\texttt{agent\_attestation} & object & Attestation evidence or reference & RECOMMENDED \\
\texttt{agent\_context\_id} & string & Identifier for the conversation/task context & RECOMMENDED \\
\bottomrule
\end{tabular}
\label{tab:capability-claims}
\end{table}

\subsection{Claim Formats and Validation}

\subsubsection{Agent Type Format}

The \texttt{agent\_type} claim uses string values from a defined taxonomy. Implementers SHOULD use one of the following standard values when applicable: \texttt{assistant} (general-purpose assistant agent), \texttt{retrieval} (information retrieval specialist), \texttt{coding} (code generation or analysis specialist), \texttt{domain\_specific} (specialized for a particular domain), \texttt{autonomous} (high degree of autonomy), or \texttt{supervised} (requiring human supervision for key actions).

Custom types MAY be used but SHOULD follow the namespaced format \texttt{vendor:type} to avoid collisions (e.g., \texttt{acme:financial\_advisor}).

\subsubsection{Delegation Chain Format}
\label{sec:delegation-chain}

The \texttt{delegation\_chain} claim is a JSON array containing objects representing each step in the delegation chain, ordered chronologically from the original user to the current agent. Each delegation step object MUST contain the following fields:

\begin{itemize}
\item \texttt{iss}: REQUIRED. String identifying the Authorization Server or entity that issued/validated this delegation step.
\item \texttt{sub}: REQUIRED. String identifying the delegator (the entity granting permission).
\item \texttt{aud}: REQUIRED. String identifying the delegatee (the agent receiving permission).
\item \texttt{delegated\_at}: REQUIRED. NumericDate representing the time the delegation occurred.
\item \texttt{scope}: REQUIRED. Space-separated string of OAuth scopes representing the permissions granted. MUST be a subset of the scopes held by the delegator.
\item \texttt{purpose}: OPTIONAL. String describing the intended purpose of this delegation step.
\item \texttt{constraints}: OPTIONAL. JSON object specifying constraints (e.g., maximum duration, allowed resources).
\item \texttt{jti}: OPTIONAL. Unique identifier for this delegation step, useful for revocation and tracking.
\end{itemize}

Relying parties MUST validate delegation chains according to the following rules: (1) verify chronological ordering based on \texttt{delegated\_at}, (2) confirm each \texttt{iss} is trusted, (3) validate that \texttt{aud} of step N matches \texttt{sub} of step N+1, (4) verify scope reduction (each step's scope is subset of delegator's scopes), (5) enforce any specified constraints, (6) validate signatures if steps are individually signed, and (7) evaluate the chain against authorization policies (e.g., maximum chain length).

\subsubsection{Agent Capabilities Format}

The \texttt{agent\_capabilities} claim is an array of string identifiers representing the agent's capabilities. Implementers SHOULD use capability identifiers from well-defined taxonomies when available. Custom capabilities SHOULD follow the namespaced format \texttt{vendor:capability}.

\subsubsection{Agent Attestation Format}

The \texttt{agent\_attestation} claim is a JSON object containing attestation evidence or a reference to it. The object MUST include a \texttt{format} field indicating the type of evidence. The recommended format is JWT-based \cite{rfc7519}, potentially compatible with IETF RATS Entity Attestation Token (EAT) \cite{ietf-rats-eat}. Other formats such as TPM 2.0 Quote \cite{tpm2} or Intel SGX Quote \cite{intel-sgx} MAY be used. An example structure:

\begin{lstlisting}[style=json]
"agent_attestation": {
  "format": "urn:ietf:params:oauth:token-type:eat",
  "token": "eyJhbGciOiJFUzI1NiIsInR5cCI6ImVhdCtqd3QifQ..."
}
\end{lstlisting}

\section{Protocol Flows}

\subsection{Agent Authentication Flow}

The OIDC-A authentication flow extends the standard OpenID Connect Authentication Code Flow \cite{openid-connect-core} with agent-specific considerations. The flow consists of four primary phases:

\textbf{Client Registration}: Clients representing agents MUST register additional metadata beyond standard OAuth 2.0 clients, as described in Section~\ref{sec:registration}. This metadata includes agent provider information, supported models, capabilities, and attestation formats.

\textbf{Authentication Request}: When initiating authentication, agents SHOULD include the \texttt{agent} scope in the authorization request to signal that agent-specific claims are requested. Agents MAY include a \texttt{delegation\_context} parameter to provide context about the delegation scenario.

\textbf{Authentication Response}: Upon successful authentication and authorization, the Authorization Server includes agent-specific claims in the ID Token. These claims include the core agent identity claims, delegation information if applicable, and attestation evidence if supported.

\textbf{Token Validation}: Relying parties MUST validate both standard OpenID Connect claims and relevant agent-specific claims. This includes validating attestation evidence, verifying delegation chains, and enforcing any delegation constraints according to local policies.

\subsection{Delegation Flow}

When authority is delegated to an agent, the following process occurs:

First, the delegator (typically a human user or another agent) authenticates with the Authorization Server and explicitly authorizes the delegation. The authorization request includes information about the agent receiving the delegation, the scope of permissions being granted, and any constraints to be imposed.

Second, the Authorization Server validates the delegation request, ensuring the delegator has the authority to grant the requested permissions and that any specified constraints are valid.

Third, the Authorization Server issues a new ID Token to the agent. This token includes the \texttt{delegator\_sub} claim identifying who granted the delegation, an updated \texttt{delegation\_chain} reflecting the new delegation step, the \texttt{delegation\_purpose}, and a \texttt{scope} claim that is appropriately constrained based on the delegation.

\subsection{Attestation Verification Flow}

To verify an agent's attestation, the following process is followed:

The agent includes attestation evidence in the \texttt{agent\_attestation} claim of its ID Token, or provides evidence through a separate mechanism. The relying party validates the evidence based on the specified \texttt{format} field.

For JWT-based attestation tokens, validation includes: verifying cryptographic signatures using trusted public keys (obtained through the discovery mechanism described in Section~\ref{sec:discovery}), comparing platform measurements against known-good reference values, validating nonces to prevent replay attacks, and checking attestation timestamps for freshness.

Optionally, relying parties MAY use the \texttt{agent\_attestation\_endpoint} (described in Section~\ref{sec:attestation-endpoint}) for validation assistance. This endpoint can provide verification services, particularly useful for complex attestation formats or when the relying party does not wish to maintain local attestation verification infrastructure.

Authorization decisions then incorporate the attestation verification status, potentially denying access to agents with unverified or failed attestation.

\section{Client Registration and Discovery}

\subsection{Agent Client Registration Metadata}
\label{sec:registration}

OIDC-A extends OAuth 2.0 Dynamic Client Registration \cite{rfc7591} with additional metadata specific to agents. Table~\ref{tab:registration} presents the agent-specific registration parameters.

\begin{table}[h]
\caption{Agent Client Registration Parameters}
\centering
\begin{tabular}{p{5cm}p{2cm}p{6.5cm}}
\toprule
\textbf{Parameter} & \textbf{Type} & \textbf{Description} \\
\midrule
\texttt{agent\_provider} & string & Identifier of the agent provider \\
\texttt{agent\_models\_supported} & array & List of supported agent models \\
\texttt{agent\_capabilities} & array & List of agent capabilities \\
\texttt{attestation\_formats\_supported} & array & List of supported attestation formats \\
\texttt{delegation\_methods\_supported} & array & List of supported delegation methods \\
\bottomrule
\end{tabular}
\label{tab:registration}
\end{table}

\subsection{Discovery Metadata}
\label{sec:discovery}

OIDC-A extends OpenID Connect Discovery 1.0 \cite{openid-connect-discovery} with additional metadata to advertise agent-specific capabilities of the Authorization Server. Table~\ref{tab:discovery} presents the additional discovery parameters.

\begin{table}[h]
\caption{Agent Discovery Parameters}
\centering
\begin{tabular}{p{5.5cm}p{1.5cm}p{6.5cm}}
\toprule
\textbf{Parameter} & \textbf{Type} & \textbf{Description} \\
\midrule
\texttt{agent\_attestation\_endpoint} & string & URL of the attestation endpoint \\
\texttt{agent\_capabilities\_endpoint} & string & URL of capabilities discovery \\
\texttt{agent\_claims\_supported} & array & Supported agent claims \\
\texttt{agent\_types\_supported} & array & Supported agent types \\
\texttt{delegation\_methods\_supported} & array & Supported delegation methods \\
\texttt{attestation\_formats\_supported} & array & Supported attestation formats \\
\texttt{attestation\_verification\_keys\_endpoint} & string & URL to retrieve verification keys \\
\bottomrule
\end{tabular}
\label{tab:discovery}
\end{table}

\section{Endpoints}

\subsection{Agent Attestation Endpoint}
\label{sec:attestation-endpoint}

The agent attestation endpoint is an OAuth 2.0 protected resource that provides attestation information about an agent or assists in validating provided attestation evidence. The endpoint URL is advertised via the \texttt{agent\_attestation\_endpoint} discovery parameter.

Request parameters include \texttt{agent\_id} (identifier of the agent being attested) and \texttt{nonce} (cryptographic nonce for freshness). The endpoint returns a JSON response containing verification status, provider information, model details, version, attestation timestamp, and cryptographic signature.

\subsection{Agent Capabilities Endpoint}

The agent capabilities endpoint provides detailed information about agent capabilities, supported constraints, and operational characteristics. The endpoint URL is advertised via the \texttt{agent\_capabilities\_endpoint} discovery parameter.

The endpoint returns a JSON response containing an array of capability objects (each with an identifier and description) and a list of supported constraints that can be applied during delegation.

\section{Security Considerations}

\subsection{Agent Authentication}

Agents SHOULD use strong, asymmetric authentication methods such as JWT Client Authentication \cite{rfc7523} or Mutual TLS \cite{rfc8705}, potentially combined with attestation evidence. Shared secret-based authentication is NOT RECOMMENDED for agents due to the increased risk of compromise in automated systems. These recommendations align with OAuth 2.0 Security Best Current Practice \cite{oauth-security-bcp}.

\subsection{Delegation Security}

Systems implementing OIDC-A MUST validate the entire delegation chain according to the rules specified in Section~\ref{sec:delegation-chain}. Each delegation step MUST enforce scope reduction, ensuring that delegated permissions are never greater than those held by the delegator. Consent mechanisms MUST be implemented to ensure delegators explicitly authorize delegation. Time-bounding of delegations is RECOMMENDED to limit the window of exposure. Policies MAY limit the maximum length of delegation chains to reduce complexity and risk. Robust revocation mechanisms are essential to terminate delegations when necessary.

\subsection{Attestation Security}

Attestation security requires multiple considerations. Signing keys used for attestation MUST be securely managed and protected from compromise. Nonce handling MUST be robust to prevent replay attacks. Known-good measurements used for comparison MUST come from trustworthy sources and be kept up-to-date. Attestation endpoints MUST be secured and rate-limited to prevent abuse. Privacy implications of attestation evidence MUST be considered, as attestation may reveal sensitive information about the agent's infrastructure or operational environment.

\subsection{Token Security}

ID Tokens containing agent claims SHOULD be encrypted \cite{rfc7516} when transmitted over potentially untrusted channels to protect sensitive agent information. Access tokens SHOULD have limited lifetimes to reduce the window of exposure if compromised. Token introspection \cite{rfc7662} MAY be used by resource servers to validate token status in real-time. The use of refresh tokens for agents requires careful consideration, as long-lived credentials increase security risks in automated systems. All tokens MUST be properly signed \cite{rfc7515} to ensure integrity and authenticity.

\section{Privacy Considerations}

Implementations of OIDC-A MUST carefully consider privacy implications. Agent identity claims may enable correlation across different relying parties, potentially allowing tracking of agent activity. Delegation chains reveal information about user-agent relationships and may expose organizational structures or workflows. User consent MUST be obtained before delegating authority to agents, and users should have visibility into how agents are using delegated permissions. Data minimization principles should be applied to agent claims, including only information necessary for authorization decisions.

\section{Compatibility and Versioning}

OIDC-A 1.0 is designed for compatibility with existing OAuth 2.0 \cite{rfc6749}, OpenID Connect Core 1.0 \cite{openid-connect-core}, JSON Web Token \cite{rfc7519}, JSON Web Signature \cite{rfc7515}, JSON Web Encryption \cite{rfc7516}, and related specifications. The extension is backward compatible in that systems not implementing OIDC-A can still interact with OIDC-A compliant systems using standard OpenID Connect flows, though without the agent-specific features. Future versions of OIDC-A will aim for backward compatibility with version 1.0, following semantic versioning principles.

\section{Conclusion}

OpenID Connect for Agents (OIDC-A) 1.0 provides a comprehensive, standards-based framework for agent identity and authorization in the OAuth 2.0 ecosystem. By extending OpenID Connect with agent-specific claims, attestation mechanisms, delegation chain support, and dedicated endpoints, OIDC-A enables secure and trustworthy integration of LLM-based agents into distributed systems.

The specification addresses the unique challenges posed by autonomous agents while maintaining compatibility with existing infrastructure. As AI agents become increasingly prevalent in digital ecosystems, standardized protocols like OIDC-A will be essential for establishing trust, managing authorization, and ensuring accountability in agent-mediated interactions. The framework aligns with modern security principles including Zero Trust Architecture \cite{nist-zero-trust}, where verification and validation occur at every interaction point.

Future work may extend OIDC-A with additional capabilities such as standardized agent capability taxonomies, enhanced privacy-preserving attestation mechanisms, integration with emerging AI governance frameworks, and support for multi-agent coordination scenarios.

\section*{Acknowledgments}

This specification builds upon the foundational work of the OAuth and OpenID Connect communities. The author thanks the broader identity and AI communities for their ongoing discussions and feedback on agent authentication and authorization challenges.

\bibliographystyle{unsrt}
\bibliography{references}

\appendix

\section{Example ID Token with Agent Claims}

The following example demonstrates a complete ID Token issued for an agent acting as an email management assistant. The token includes core agent identity claims, delegation information, attestation evidence, and capability declarations.

\begin{lstlisting}[style=json, caption=Complete Agent ID Token Example]
{
  "iss": "https://auth.example.com",
  "sub": "agent_instance_789",
  "aud": "client_123",
  "exp": 1714435200,
  "iat": 1714348800,
  "auth_time": 1714348800,
  "nonce": "n-0S6_WzA2Mj",
  "agent_type": "assistant",
  "agent_model": "gpt-4",
  "agent_version": "2025-03",
  "agent_provider": "openai.com",
  "agent_instance_id": "agent_instance_789",
  "delegator_sub": "user_456",
  "delegation_purpose": "Email management assistant",
  "agent_capabilities": [
    "email:read",
    "email:draft",
    "calendar:view"
  ],
  "agent_trust_level": "verified",
  "agent_context_id": "conversation_123",
  "agent_attestation": {
    "format": "urn:ietf:params:oauth:token-type:eat",
    "token": "eyJhbGciOiJSUzI1NiIsInR5cCI6IkpXVCJ9...",
    "timestamp": 1714348800
  },
  "delegation_chain": [
    {
      "iss": "https://auth.example.com",
      "sub": "user_456",
      "aud": "agent_instance_789",
      "delegated_at": 1714348700,
      "scope": "email profile calendar"
    }
  ]
}
\end{lstlisting}

This example illustrates several key aspects of OIDC-A: (1) Standard OpenID Connect claims (\texttt{iss}, \texttt{sub}, \texttt{aud}, \texttt{exp}, \texttt{iat}) establish the token's validity and scope. (2) Core agent claims identify the specific agent type, model, and provider. (3) The delegation chain shows a single-step delegation from user\_456 to the agent. (4) Agent capabilities are explicitly declared, enabling fine-grained authorization decisions. (5) Attestation evidence is included in JWT format for verification.

\section{Example Multi-Step Delegation Chain}

Complex agent workflows often involve multiple delegation steps, where one agent delegates a subset of its authority to another specialized agent. The following example demonstrates a two-step delegation chain where a user's email management agent delegates calendar viewing permission to a scheduling analysis agent.

\begin{lstlisting}[style=json, caption=Multi-Step Delegation Chain Example]
{
  "delegation_chain": [
    {
      "iss": "https://auth.example.com",
      "sub": "user_456",
      "aud": "agent_instance_789",
      "delegated_at": 1714348800,
      "scope": "email calendar",
      "purpose": "Manage my emails and calendar"
    },
    {
      "iss": "https://auth.example.com",
      "sub": "agent_instance_789",
      "aud": "agent_instance_101",
      "delegated_at": 1714348830,
      "scope": "calendar:view",
      "purpose": "Analyze available time slots"
    }
  ]
}
\end{lstlisting}

This example demonstrates several important properties of delegation chains: (1) \textbf{Chronological ordering}: The \texttt{delegated\_at} timestamps show the temporal sequence of delegations. (2) \textbf{Scope reduction}: The second delegation step (\texttt{calendar:view}) is a strict subset of the first step's scope (\texttt{email calendar}). (3) \textbf{Audience chaining}: The \texttt{aud} of the first step (\texttt{agent\_instance\_789}) matches the \texttt{sub} of the second step, establishing the chain of authority. (4) \textbf{Purpose tracking}: Each step includes a \texttt{purpose} field documenting the intent of the delegation. (5) \textbf{Issuer consistency}: All delegation steps are validated by the same trusted Authorization Server.

Relying parties receiving this delegation chain would validate that: the original user (\texttt{user\_456}) authorized the initial delegation; the intermediate agent (\texttt{agent\_instance\_789}) had authority to further delegate; the scope was appropriately reduced at each step; all issuers are trusted; and the chain length complies with local policies.

\section{Implementation Considerations}

\subsection{Deployment Scenarios}

OIDC-A can be deployed in various architectural patterns:

\textbf{Centralized Pattern}: A single Authorization Server manages all agent identities and delegations. This pattern simplifies trust management but may create a single point of failure.

\textbf{Federated Pattern}: Multiple Authorization Servers operate in a federated trust model, allowing agents from different providers to interact. This requires careful trust establishment between Authorization Servers.

\textbf{Hybrid Pattern}: Some agents are managed by a central Authorization Server while others are managed by provider-specific servers with federation relationships.

\subsection{Performance Considerations}

Implementing OIDC-A requires consideration of performance impacts:

\textbf{Token Size}: Delegation chains and attestation evidence can significantly increase ID Token size. Implementations should consider using references to external resources when tokens become too large, following best practices from OAuth 2.0 for Native Apps \cite{rfc8252} regarding token handling.

\textbf{Validation Overhead}: Validating delegation chains and attestation evidence adds computational overhead. Caching validation results where appropriate can improve performance.

\textbf{Attestation Freshness}: Frequent attestation verification may impact performance. Implementations should balance security requirements with performance constraints.

\end{document}